\documentclass[singlecolumn,superscriptaddress,amsmath,amssymb,floatfix,aps,showpacs]{revtex4-1}

\usepackage{amsmath}
\usepackage{graphicx}
\usepackage{hyperref}

\usepackage{color}

\usepackage{graphicx}
\usepackage[usenames,dvipsnames,svgnames,table]{xcolor}
\newcommand{\reva}[1]{{\color{black}  #1}}
\newcommand{\revb}[1]{{\color{black}  #1}}

\begin{document}


\title{The power of being explicit: demystifying work, heat, and free energy in the physics of computation} 

\author{Thomas E. Ouldridge$^1$,*  Rory A. Brittain$^2$, Pieter Rein ten Wolde$^3$
\\{\small $^1$ Department of Bioengineering and Centre for Synthetic Biology, Imperial College London, London, SW7 2AZ, UK; t.ouldridge@imperial.ac.uk}
\\{\small $^2$ Department of Mathematics, Imperial College London, London, SW7 2AZ, UK; r.brittain15@imperial.ac.uk}
\\{\small $^3$ FOM Institute AMOLF, Science Park 104, 1098 XE Amsterdam, The Netherlands; tenwolde@amolf.nl}
\\ {\small *To whom correspondence should be addressed.}}



\maketitle


\section{introduction}
Interest in the thermodynamics of computation has revived in recent years, driven by developments in science, economics and technology. Given the consequences of the growing demand for computational power, the idea of reducing the energy cost of computations has gained new importance (\cite{DeBenedicts2017,Frank2017}). Simultaneously, many biological networks are now interpreted as information-processing or computational systems constrained by their underlying thermodynamics (\cite{Andrieux2008,Lan2012,Mehta2012,Ito2015,govern2014,Barato2014,Mehta2016,tenWolde2016,Ouldridge_polymer_2017,Ouldridge_comput_2017,Barato2017}). Indeed, some suggest (\cite{Bennett1982,Adleman1994,Organick2018}) that low-cost, high-density biological systems may help to mitigate the rising demand for computational power and the ``end" of Moore's law of exponential growth in the density of transistors (\cite{Frank2017}). 

In this chapter we address  widespread misconceptions about thermodynamics and the thermodynamics of computation. \revb{In particular, we will argue against the general perception that a measurement or copy operation can be performed at no cost; against the emphasis placed on the significance of erasure operations; and against the careless discussion of heat and work.} While not universal, these misconceptions are sufficiently prevalent (particularly within interdisciplinary contexts) to warrant a detailed discussion. In the process, we will argue that explicitly representing fundamental processes is a useful tool, serving to demystify key concepts.

We  first give a brief overview of thermodynamics, then the history of the thermodynamics of computation - particularly in terms of copy and measurement operations inherent to classic thought experiments. Subsequently, we analyse these  ideas via an explicit biochemical representation of the entire cycle of Szilard's engine. In doing so we show that molecular computation is both a promising engineering paradigm, and a valuable tool in providing fundamental understanding.

\section{Basic thermodynamics}
\label{sec:thermo}
Here we briefly recap the key ideas of work, heat, entropy and free energy. In effect, we introduce several consistent ways to think about the second law of thermodynamics. Experienced readers may wish to skip to Section~\ref{sec:demon}; those wishing for more detail on the background of equilibrium thermodynamics and statistical mechanics may refer to \cite{Atkins2010} and \cite{Huang1987}. Note that the quantities we consider here are implicitly statistical averages, rather than quantities defined for individual fluctuating trajectories (\cite{Jarzynski1997,Crooks1999,Esposito2011,Seifert2005}), and that we will restrict ourselves to the classical regime. 

\subsection{Work, heat, the second law and entropy}
\label{sec:2nd law}

Energy is conserved, but  not all energy is equivalent. Certain forms of energy, like that stored when a mass is raised in a gravitational field, are more useful than others, such as that  stored in a hot object. The energy of an ideal mass is stored in one accessible (collective) degree of freedom: the position of the centre of mass in a gravitational field. The energy of a hot object such as a volume of gas is stored in many degrees of freedom (the momenta of the gas particles). Unlike the mass, it is impossible for an experimenter to couple to these degrees of freedom individually and use the stored energy. This distinction underlies the second law. 

Energy is transferred as heat when the transfer occurs through many degrees of freedom that are individually inaccessible. By contrast, work is done when the transfer of energy occurs via the accessible degrees of freedom. We define a heat reservoir as an ideal, arbitrarily large system that can only accept energy transfer via heat (think of a large gas at fixed volume). The heat reservoir is characterised by a temperature $T$, which quantifies the average energy per degree of freedom. We define an ideal work reservoir as an ideal system, like a mass in a gravitational field, that only stores energy in an accessible degree of freedom.  \revb{We note that this definition of work -- in terms of the transfer of energy via the manipulation of accessible coordinates -- is consistent with the definition used in modern stochastic thermodynamics (\cite{Horowitz2008}). The definition of heat is then, necessarily, also consistent.}

\begin{figure}
\begin{center}
  \includegraphics[width=0.8\textwidth]{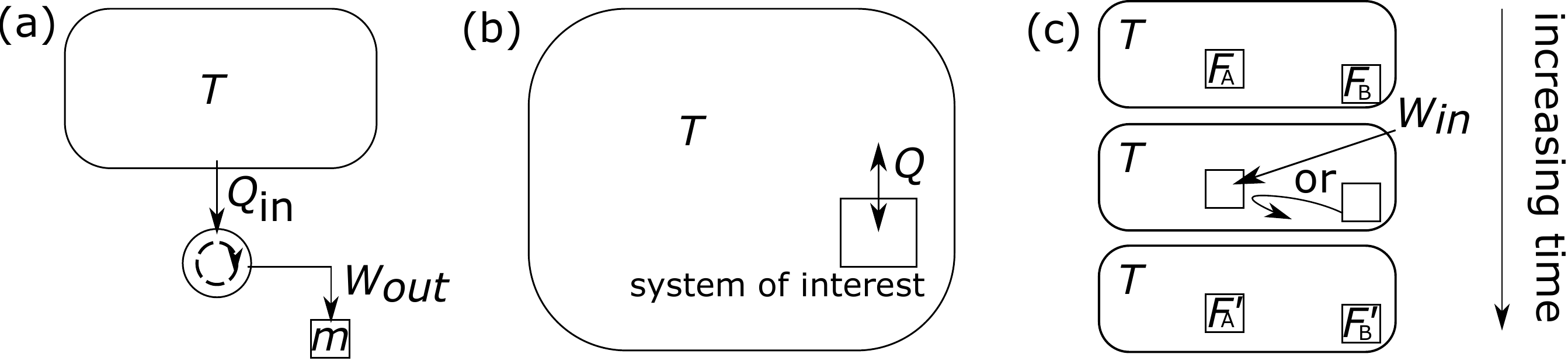}
\caption{(a) Graphical representation of processes forbidden by the second law of thermodynamics: a cyclic engine cannot soley take in heat from a reservoir $T$ and convert it to work to raise a mass $m$. (b) Illustration of the canonical ensemble: a small system of interest is embedded within a larger system setting the temperature $T$. Heat $\mathcal{Q}$ can be exchanged between system and reservoir. (c) The input of external work, or the coupling to another subsytem, can be used to drive up the free energy of a subsystem $A$ so that $\mathcal{F}_{A}^\prime > \mathcal{F}_{A}$. In the absence of the seconf subsystem, the second law dictates that $\mathcal{W}_{\rm in}\geq  \mathcal{F}_{A}^\prime - \mathcal{F}_{A}$. In the other case of no external work input, $\mathcal{F}_B - \mathcal{F}_B^\prime \geq  \mathcal{F}_A^\prime - \mathcal{F}_A$.}
\label{fig:heat_work}       
\end{center}
\end{figure}

One of the traditional statements of the second law is:
"No process is possible whose sole result is the conversion of heat into work."
This statement makes precise the idea that ``not all energies are equal". No engine exists that cyclically takes energy from a single heat reservoir and deposits all of it into a work reservoir (Fig.~\ref{fig:heat_work}\,(a)). The cyclical operation is important here -- if the engine underwent a sustained change, then the transfer of energy would not be the "sole result" of the  process.

A more typical modern statement of the second law would be:
"In a closed system, the entropy is non-decreasing."
The entropy quantifies the uncertainty in the precise state of a system. Specifically (\cite{Jaynes1957}),
\begin{equation}
\mathcal{S}[p(x)] = -k_{\rm B} \sum_x p(x) \ln p(x).
\label{eq:entropy}
\end{equation}
Here, $x$ denotes the possible states of a system, and $p(x)$ the probability that the system will be found in state $x$. The flatter the distribution $p(x)$, the larger the entropy. If we are completely certain of the state of a system, then all terms in the sum in Eq.~\ref{eq:entropy} are identically zero. 
 
Entropy incorporates the traditional statement of the second law. When energy is transferred to a work reservoir, there is no change in the work reservoir's entropy since it only has one degree of freedom -- its energy uniquely specifies its state. By contrast, when energy is transferred to a heat reservoir, it is shared between the many degrees of freedom. More energy increases the uncertainty, since there are more ways to divide the energy up. Thus the transfer of energy to a heat reservoir is associated with an entropy increase, and converting heat  directly into work is forbidden because it would be an overall decrease in the entropy of a closed system. To be precise, if heat ${\rm d} \mathcal{Q}$ is transferred to a heat reservoir, the reservoir's entropy increase is ${\rm d} \mathcal{S} = T{\rm d}\mathcal{Q}$. 
Importantly, the entropic definition of the second law allows us to analyse processes in which systems that are not ideal reservoirs undergo changes,  using Eq.~\ref{eq:entropy}; this framework allows us to analyse non-cyclic systems, and individual steps of cyclic operations. 

\subsection{Equilibrium, the Boltzmann distribution and free energies}
\label{sec:equilibrium}
An isolated system \revb{with sufficiently many interacting degrees of freedom will evolve towards an equilibrium steady state in which there is no systematic variation of observables with time, or net currents of matter or energy}. Since the entropy of an isolated system cannot decrease, the equilibrium distribution $p_{\rm eq}(x)$ maximises the entropy subject to whatever constraints are relevant (e.g., the overall conservation of energy). It is often convenient to consider a small system of interest coupled to a heat reservoir at temperature $T$, as illustrated in Fig.~\ref{fig:heat_work}\,(b). In this case, representing the states of the small system  by $x$, it is possible to show that
\begin{equation}
p_{\rm eq}(x) \propto \exp(-\epsilon(x)/k_{\rm B}T),
\label{eq:Bmann}
\end{equation} 
where $\epsilon(x)$ is the energy of configuration $x$ (\cite{Huang1987}). 
The distribution peq(x)  maximizes the overall entropy of the combined system and reservoir given a fixed total energy to share between the two. 
For simplicity, we restrict ourselves to this "canonical ensemble".

Prior to reaching equilibrium, systems have a non-equilibrium distribution, $p(x) \neq p_{\rm eq}(x)$. The distance from equilibrium can be characterised by the non-equilibrium generalized free energy (\cite{Esposito2011,Parrondo2015})
\begin{equation}
\begin{array}{c}
\mathcal{F}[p(x)] =  \sum_x p(x) \epsilon(x) + k_{\rm B}T \sum_x p(x)  \ln p(x) = \langle \epsilon \rangle - T\mathcal{S}[p(x)]
\\
\\
=\mathcal{F}[p_{\rm eq}(x)] + k_{\rm B}T\sum_x p_{\rm}(x) \ln \frac{p_{\rm}(x)}{p_{\rm eq}(x)} \geq  \mathcal{F}[p_{\rm eq}(x)].
\end{array}
\label{eq:free_energy}
\end{equation}
\revb{Following a process in which $p(x)$ evolves over time,} the change in the total entropy of the system and reservoir is given by $-\Delta \mathcal{F}[p(x)]/T$;  $\mathcal{F}[p(x)]$ then {\it decreases} over time, and is {\it minimal} at equilibrium. This minimum represents a balance between minimisation of energy $\langle \epsilon \rangle$ and maximisation of entropy $\mathcal{S}[p(x)]$ within the system.  



Although non-equilibrium free energies cannot increase spontaneously, they can be driven upwards as in Fig.~\ref{fig:heat_work}\,(c). Specifically, the change in free energy determines the minimal work that must be applied to move a system $A$ coupled to a heat reservoir between $p(x)$ and $p^\prime(x)$ (\cite{Esposito2011}): $\mathcal{W}_{\rm in} \geq \mathcal{F}_{A}[p^\prime(x)] - \mathcal{F}_{A}[p(x)]$. As a consequence, the maximum work that can be extracted is $\mathcal{W}_{\rm  out} \leq \mathcal{F}_{A}[p(x)] - \mathcal{F}_{A}[p^\prime(x)]$. The system's free energy must change to obtain positive work since work cannot be extracted from the heat reservoir alone.

We could also couple to another system that is not an idealised work reservoir (Fig.~\ref{fig:heat_work}\,(c)). In this case, the second law dictates that the total free energy of the combined system must decrease, but a decrease in the free energy of one subsystem can compensate for an increase in the other. 
A non-equilibrium system is then fundamentally an exploitable resource. If a system $A$ is out of equilibrium, we can in principle couple it to a second system $B$ and use the relaxation of $A$ to drive a change in $B$. If $A$ is in equilibrium, however, it is not exploitable in this way.

\reva{It is important to note that, although the free energy has the dimensions of an energy, and although changes in free energy are bounded by work input, free energies are not energies in a very fundamental sense. Energies obey the first law of thermodynamics; energy can never be destroyed, only converted between different forms. If work is done on a system, either its internal energy must change, or energy must be transferred to the environment as heat. Free energies, however, also have an entropic component (Eq. \ref{eq:free_energy}). They are therefore not conserved; it is perfectly possible for the free energy of a system to decrease with no other consequences. We shall explore this fact in more detail in Section~\ref{sec:focus}.} 

\subsection{Thermodynamic reversibility}
\label{sec:reversibility}
\reva{Consider applying a time-varying manipulation of a system, driving it from $p(x)$ to $p^\prime(x)$. If and only if this process is thermodynamically reversible, applying the manipulation in a time-reversed manner would convert $p^\prime(x)$ back to $p(x)$, and the intermediate probability distributions obtained during the forward manipulation would be reproduced in the opposite order during the reverse protocol \cite[see][]{Sagawa2014}. Thermodynamically reversible processes necessarily involve no change in the overall entropy of the system and reservoirs to which they are coupled, because a positive entropy increase on the forward manipulation would correspond to a forbidden decrease on the reverse. Reversible processes therefore require work input of exactly $\mathcal{W}_{\rm in} =F[p^\prime(x)] - F[p(x)]$ \cite{Esposito2011} and constitute an important idealized limit of practical processes. 
}

\revb{
\section{The physics of computation}
To perform computation, a physical system -- be it silicon-based or biological -- must be able to change in response to a range of inputs. Indeed, data and information must have a physical realisation simply to exist, let alone to be manipulated during a computation (\cite{Landauer1991}). If computation involves changes in the states of physical systems, then these processes can be analysed thermodynamically, and the resultant thermodynamic costs and requirements can be assessed. This approach is becoming increasingly fashionable, driven by a desire to reduce the consumption of electricity by our increasing demands on computational power (\cite{DeBenedicts2017,Frank2017}).}

\revb{Although recent years have seen a wider focus (see e.g. \cite{Owen2017}), the study of the thermodynamics of computation has been dominated by the discussion of two processes -- erasure and measurement (or copying) of single two-state memories. Partly, this focus follows from the fact tht these processes are prototypical computational operations that provide much of the framework for understanding more complex systems. Equally importantly, however, erasure and measurement lie at the heart of a fundamental debate in physics relating to the validity and meaning of the second law of thermodynamics. In the following paragraphs, we summarise this debate.}

\subsection{A history of Maxwell's demon and Szilard's engine}
\label{sec:demon}
Physicists have long been concerned by the second law. Most famously, James Clerk Maxwell wondered whether an intellegent being (a "demon") could violate the second law by extracting work from random fluctuations of an equilibrium system (\cite{Dougal2016}). Maxwell's demon operated a trap door separating two volumes of gas at initially equal temperature. If the demon could determine the velocity of gas particles, it could selectively open the door to allow high energy particles to pass in one direction, and low energy particles in the other. Over time a temperature difference would develop, to which a work-extracting heat engine could be coupled. There is no fundamental lower bound on the work needed to open and close a trap door. Therefore it appeared to Maxwell that the demon could use its intellegence to perform measurement and feedback to exploit fluctuations and thereby violate the second law of thermodynamics by extracting work from an initially equilibrated system.

Many authors have attempted to explain why the second law  is not violated by Maxwell's demon. Typically, they construct a demon-like system, analyze its behaviour and demonstrate that violation of the second law does not occur. This approach has been criticised (\cite{Norton2005}), on the grounds that the accuracy of the second law is generally assumed in setting up the analysis. We, however, believe that it is valid both to test the self-consistency of the second law in this way, and to ask the conceptual question of {\it how} the self-consistency is manifested, as the understanding gained applies beyond the original context.

\begin{figure}[!]
\begin{center}
  \includegraphics[width=0.9\textwidth]{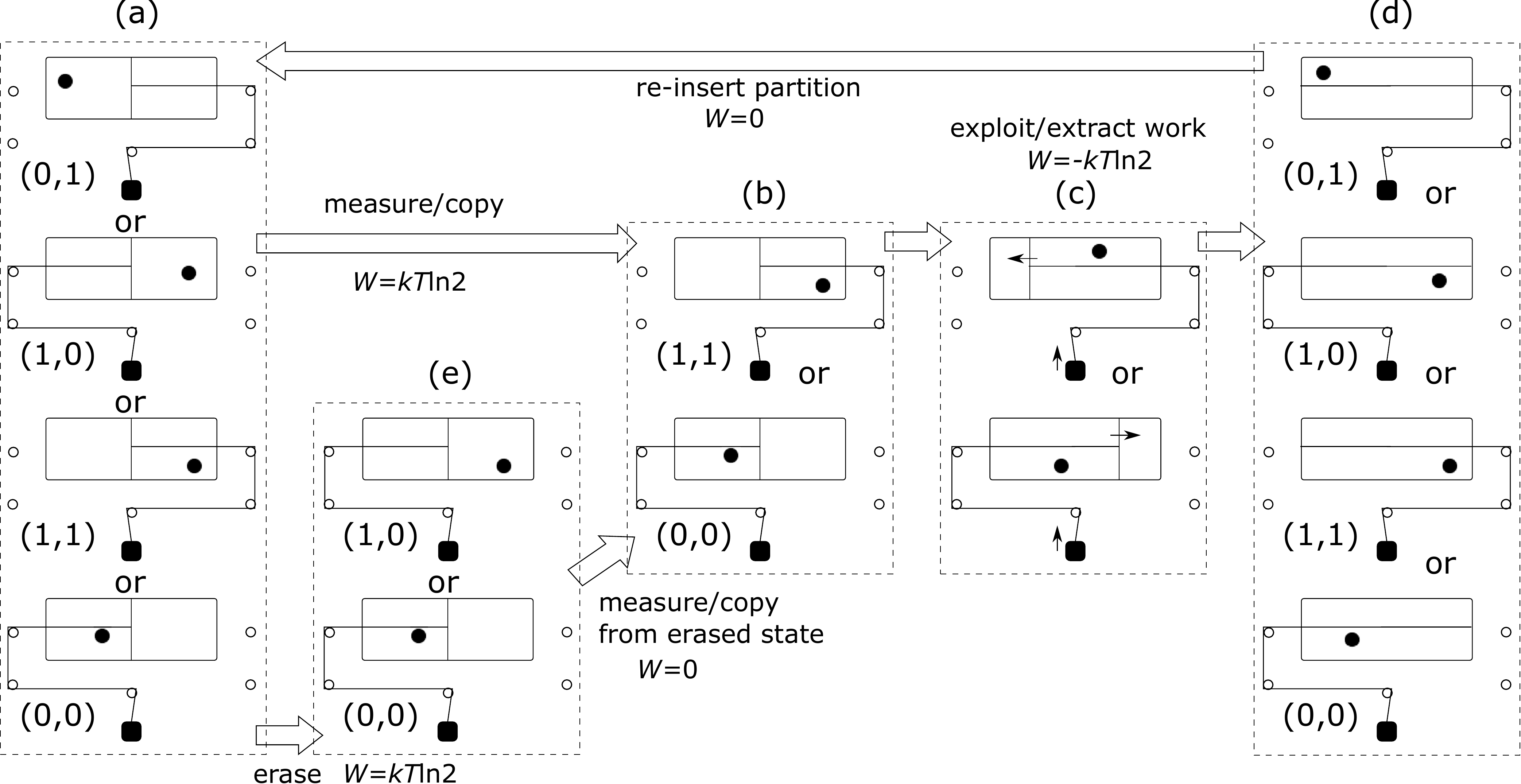}
\caption{A cycle of Szilard's engine, with possible operating states illustrated. Steps are labelled with the minimal work input required for reversible operation (negative values correspond to work extraction). The engine consists of a single particle in a volume, a partition, and a weight that can be attached to either side of the partition via pulleys. (a) Initially the particle and the weight can both be associated with either side of the volume, and four states are possible (labelled $(d,m)$, where $d=0(1)$ corresponds to the particle on the left(right) hand side of the system, and $m=0(1)$ corresponds to the apparatus on the left(right) side of the partition). (b) A measurement correlates the weight's connection with the particle's position. Two states are possible after a perfectly accurate measurement. (c) The particle is allowed to expand against the partition, raising the weight. Note that the weight's attachment state does not change. Eventually state (d) is reached, when the particle's position is uncorrelated with the weight's attachment. The system is then restored to its initial condition by re-inserting the partition. (e) Alternatively, one might first ``erase", setting the pulleys into a guaranteed configuration and leaving two possible states overall, prior to performing the measurement. }
\label{fig:Szilard}       
\end{center}
\end{figure}

Whilst the term "Maxwell's demon" is used quite liberally, we will focus on systems in which a measurement is made, stored, and then used to implement a measurement-dependent feedback.
In 1929, Leo Szilard described an analytically tractable "engine" with measurement and feedback (\cite{Szilard1964}). A version is schematically illustrated in Fig.~\ref{fig:Szilard}. Here, the demon simply determines whether a single particle is located on the left or right hand side of a piston in a hollow cylinder, and configures the system to exploit this fact. Given knowledge of the particle's position, the piston can be coupled to a weight so that the pressure of the particle can do work to raise the weight. It is clear that measurement and feedback are required to couple the weight correctly. Assuming that this procedure is performed with 100\% accuracy, the maximum work extracted per expansion can be calculated by integrating $p {\rm d}V$. Here, $p$ is the average pressure applied by the particle and $V$ is the volume in which it is contained, which grows from $V_{\rm cylinder}/2$ to $V_{\rm cylinder}$. The result is $k_{\rm B}T \ln 2$.

Szilard sought to identify a cost that compensated for the apparent ability to extract work from the initial equilibrium system. He realised that the essence of his system could be analysed without reference to an intellegent being. Instead, he argued that the demon's measurement generates a correlation between two physical degrees of freedom. In Fig.~\ref{fig:Szilard}, these degrees of freedom  are the position of the particle (the data) and the weight's connection to the partition (the memory). Crucially, Szilard emphasized that post-measurement, the memory must retain its state even if the data subsequently changes. In Fig.~\ref{fig:Szilard}, for example, the weight must stay attached to the same side of the piston even as the particle's accessible volume expands to fill the whole cylinder. 
Szilard  stated that "if the measurement could take place without compensation", then the second law could be violated.  He claimed that there must be a minimum "production of entropy" during measurement that compensated for the $k_{\rm B}T \ln 2$ of work extracted during the expansion. \revb{In the computational sense, he was arguing that there is a cost to copying of data}.


In 1951, Brillouin reopened the question. He too believed that the demon needed to pay a cost when measuring (\cite{Brillouin1951}). He associated this cost with the generation of a non-equilibrium photon that could be used to  see the particle. Brillouin did not consider correlating a memory with the measured system, but merely the interaction mechanism. His proposed measurement costs are therefore of a fundamentally different character from Szilard's.

A decade later, Landauer asked whether heat generation was necessary during computation (\cite{Landauer1961}). Although he referred to Brillouin's work, he did not deal with demons. Rather, he argued that computational operations without a single-valued inverse (many-to-one functions) necessarily lead to an increase in the entropy of the environment. Landauer pointed out that such "logically irreversible" functions -- for example, "erasing" a bit of unknown value to zero - decrease the  entropy of the bit itself. He argued that there must be an entropy increase in the environment to preserve the second law. For a symmetric bit initially in state 0 with 50\% probability, the initial entropy is $k_{\rm B} \ln2$ (from Eq.~\ref{eq:entropy}). After erasure is complete, the entropy of the bit is 0. Thus the entropy of the environment must increase by at least $k_{\rm B} \ln 2$ to compensate. In a simple system with single heat and work reservoirs, this operation would involve a transfer of at least $k_B T\ln 2$ of energy from the work reservoir to the heat reservoir via the bit.

Eventually, Bennett drew upon Landauer's work to argue that measurement, \revb{or copying of a bit}, could be performed with no work input, provided that the memory was in a well-defined initial state (say, 0) (\cite{Bennett1982, Bennett2003}). In this case, the measurement is logically reversible, since two initial states of the memory and data are mapped to two final states, as shown in Fig.~\ref{fig:Szilard}. Such a measurement requires no net increase in the entropy of the environment, and hence no work input. However, to operate again the memory must be reset. If the "data" bit has already been exploited at this point, for example by a demon extracting work, it no longer retains its previously measured value.  The reset cannot then be performed in a logically reversible manner -- we must perform a logically irreversible "erase" (Fig.~\ref{fig:Szilard}) to reset the memory. This reset must increase the entropy of the environment by Landauer's principle, and hence the work extracted ($\mathcal{W}_{\rm out} \leq k_{\rm B}T \ln2$) is paid for by the resetting of the device ($\mathcal{W}_{\rm in} \geq k_{\rm B}T \ln2$). Bennett argued that this erasure cost explains why the second law is not violated. 

Within the wider scientific community, Bennett's argument is generally accepted as the ``true" explanation of why the second law survives (e.g. \cite{feynman1998feynman,Plenio2001,Bub2001,Dillenschneider2009,Maruyama2009,berut2012,Jun2014}). Although there are exceptions to this perspective (\cite{Fahn1996,Maroney2005,Sagawa2009,Sagawa2014,Parrondo2015}), Szilard's argument  that "measurement cannot take place without compensation" has largely been discounted, with measurement viewed as potentially costless, and the cost of erasure viewed as fundamental to resolving the paradox.  Here, we argue against this perspective, leveraging a concrete molecular mechanism for implementing the entire measurement and feedback cycle of Szilard's engine. We  use this mechanism to demonstrate that, in fact, Szilard's emphasis on the cost of correlating degrees of freedom, which corresponds to creating a non-equilibrium state, is essentially correct. We will  simultaneously highlight fundamental issues that follow from an over-emphasis on the significance of erasure, and the stages at which work and heat are exchanged with the environment.

Our explicit molecular mechanism overcomes the inherent inscrutability of Maxwell's demon. Previous work has included both physical and thought experiments on systems that perform erasure, measurement and exploitation (\cite{Landauer1961,Szilard1964,Bennett1982,Sagawa2009,Lambson2011,berut2012,Jun2014,Koski2014,Hong2016,Ouldridge_comput_2017}). However, it is extremely rare to study a full cycle of Szilard's engine in which all components and information-processing mechanisms are concrete and explicitly accounted for (we leverage our recent work in \cite{Brittain2018}). Based on our analysis of a molecular Szilard engine, we will make the case for explicit molecular systems as a natural environment in which to embed thermodynamic thought experiments more generally.




\section{Implementing Szilard's engine with molecular reactions}
\label{sec:molecular demon}
Well-mixed, dilute molecular systems fit naturally into the paradigm of Section~\ref{sec:thermo} (\cite{Ouldridge_review_2018}). 
The state of such systems is given by the number of each molecular species present. Changes in state are chemical reactions such as $A+B \rightarrow C+D$, which happen at rates determined by reactant concentrations. Some of these reactions may correspond to large-scale conformational changes of a single molecule. ``Chemical Reaction Networks" (CRNs) of this kind have been extensively studied in the biophysics, mathematics and computer science literature (\cite{Krishnamurthy2017}), and provide a realistic model of many biological processes. Furthermore, it has been shown that arbitrary CRNs can be approximated in the laboratory through nucleic acid-based designs (\cite{Qian2011b,Chen2013,Srinivas2017,Plesa2018}).


The discrete nature of the state space at the CRN level introduces a minor modification to the theory of Section~\ref{sec:thermo}. These "macrostates" each contain an enormous number of microscopic configurations.
Macrostates are then stabilised not only by having low energy, but also by having many microscopic configurations (an "entropic" contribution).  Thankfully, we can simply work at the level of macrostates, and replace the energies of individual states $\epsilon(x)$ appearing in Eqs.~\ref{eq:Bmann}-\ref{eq:free_energy} with chemical free energies of macrostates $f_{\rm chem}(m)$, which incorporate both energetic and entropic contributions.
\begin{equation}
\begin{array}{c}
f_{\rm chem}(m) = k_{\rm B}T \ln \sum_{x \in m} \exp(-\epsilon(x)/k_{\rm B}T),
\hspace{5mm}
p_{\rm eq} (m) \propto \exp(-f_{\rm chem}(m)/k_{\rm B}T),
\vspace{2mm}
\\
\mathcal{F}[p(m)] = \sum_m p(m)f_{\rm chem}(m) + k_{\rm B}T \sum_m p(m)  \ln p(m) = \langle f_{\rm chem} \rangle - T\mathcal{S}[p(m)]
\vspace{2mm}
\\
=\mathcal{F}[p_{\rm eq}(m)] + k_{\rm B}T\sum_m p_{\rm}(m) \ln \frac{p_{\rm}(m)}{p_{\rm eq}(m)} \geq  \mathcal{F}[p_{\rm eq}(m)].
\end{array}
\label{eq:free_energy_m}
\end{equation}
For a detailed discussion of the macrostate perspective, see \cite{Ouldridge_review_2018}.

\begin{figure}
\begin{center}
  \includegraphics[width=0.7\textwidth]{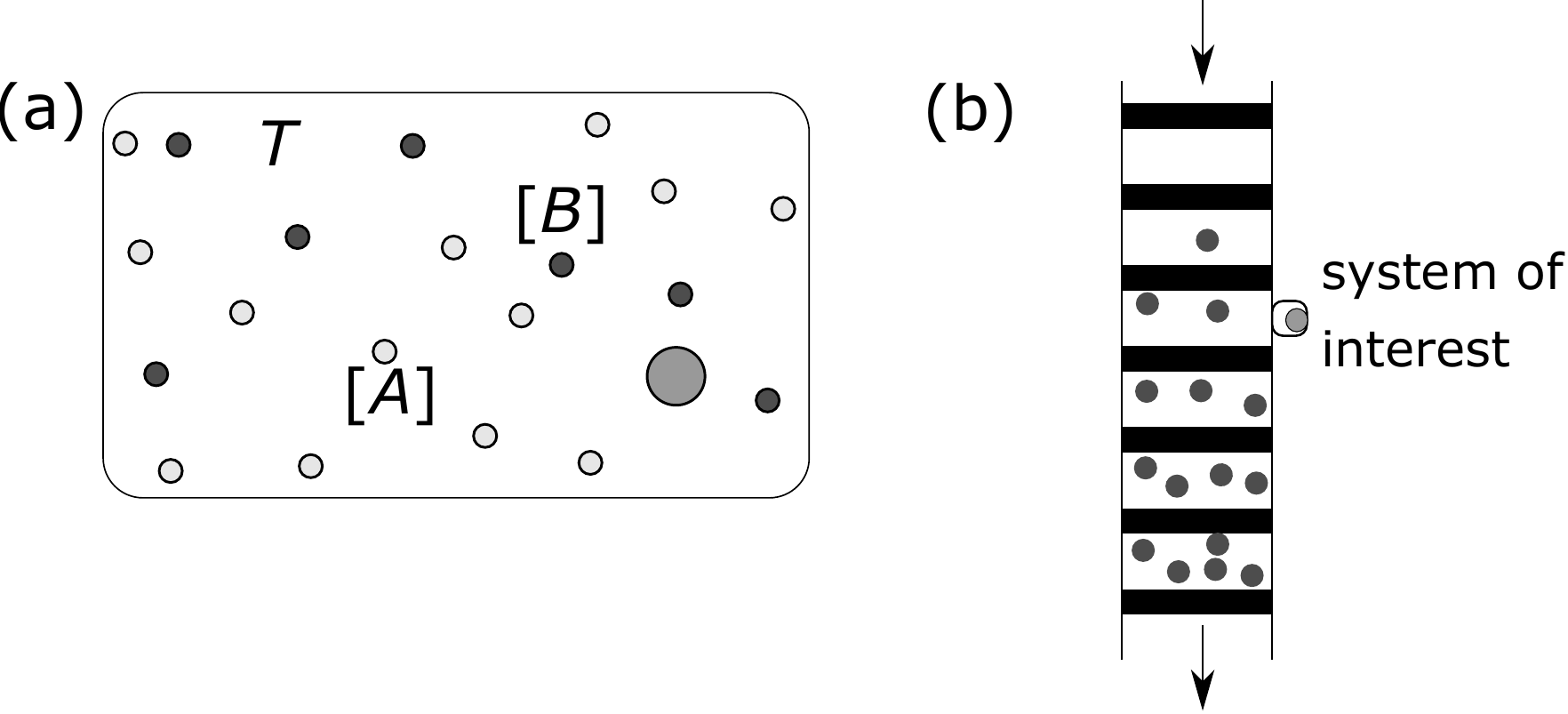}
\caption{Molecular Buffers. (a) A single molecule of interest in a large bath of fuel molecules of two types, $A$ and $B$, at a temperature $T$ and concentrations $[A]$ amd $[B]$, respectively. (b) A system of interest coupled to one of a series of buffers at different concentrations of fuel molecles.}
\label{fig:baths}       
\end{center}
\end{figure}

We will consider large baths of "fuel" molecules, or buffers;
our systems will inter-convert these fuel molecules (\cite{Seifert2011}). The change in chemical free energy of a system when an additional molecule of $A$ is added is called the {\it chemical potential} of $A$, $\mu_A$. For dilute systems (\cite{Ouldridge_review_2018,Nelson2004}), 
\begin{equation}
\mu_{A} = \mu^0_{A} + k_{\rm B}T \ln \left( [A]/[1{\rm M}] \right),
\end{equation}
where $\mu^0_A$ is the (theoretical) value of the chemical potential at a reference concentration of 1M, and $[A]$ is its actual concentration. If a reaction consumes a single $A$, and produces a single $B$, then there is a contribution to the overall chemical free energy change of
\begin{equation}
\Delta f_{\rm chem}(A \rightarrow B) = \mu_B - \mu_A = \mu^0_B - \mu^0_A + k_{\rm B}T \ln \left( [A]/[B] \right).
\end{equation}
This contribution to $\Delta f_{\rm chem}$ can be used to drive the interconversion of other species. For example, consider the reaction $A+C \rightleftharpoons B +D$.  A large, negative value of $\Delta f_{\rm chem}(A \rightarrow B)$ would tend to favour the conversion of $C$ into $D$, because of the more favorable chemical free energy of B. 

\subsection{Buffers as a source of chemical work}
\label{sec:buffers}
We will assume that the buffers are sufficiently large that any reactions have a negligible effect on the probability distribution of bath macrostates. In this limit, the baths become reservoirs of "chemical work", analogous to the (less concrete) work reservoirs introduced in Section~\ref{sec:2nd law}. In this case, we now use $m$ to label the macrostate of only the molecules that are not part of the buffers (the "system of interest"), with $p(m)$ describing the distribution of these macrostates. The change in free energy due to an arbitrary process is then
\begin{equation}
\begin{array}{c}
\Delta\mathcal{F} = \Delta \left(\sum_m p(m)f^{\rm sys}_{\rm chem}(m) \right) + k_{\rm B}T \Delta \left(\sum_m p(m)  \ln p(m) \right) + \langle \sum_n \Delta f^n_{\rm chem} \rangle
\vspace{2mm}
\\
= \Delta \langle f^{\rm sys}_{\rm chem} \rangle - T\Delta \mathcal{S}^{\rm sys}[p(m)] + \langle \Delta f_{\rm chem}^{\rm buffers} \rangle = \Delta\mathcal{F}^{\rm sys} - \mathcal{W}_{\rm chem},
\end{array}
\label{eq:overall_change}
\end{equation} 
where $f^{\rm sys}_{\rm chem}(m)$ is the chemical  free energy of the system of interest only, and $ \langle \Delta f^n_{\rm chem} \rangle$ is the expected change in chemical free energy of buffer $n$. The chemical work is defined as, $- \langle \Delta f_{\rm chem}^{\rm buffers} \rangle =-\langle \sum_n \Delta f^n_{\rm chem} \rangle = \mathcal{W}_{\rm chem}$; a positive value allows an increase in the free energy of the system of interest whilst satisfying $\Delta \mathcal{F} \leq0$ overall.

By varying the chemical potentials of fuel molecules, we can implement a time-varying protocol on the system of interest (\cite{Ouldridge_comput_2017,Rao2016,Schmiedl2007,Rao2018}). 
A device for implementing such a protocol is illustrated in Fig.~\ref{fig:baths}\,(b). We have a few molecules of interest that are trapped inside a small reaction volume. These molecules could be tethered to the surface of the volume, or trapped by a membrane. Outside of the volume is a large bath of fuel molecules, which can diffuse into the volume, setting the chemical potential and biasing reactions as desired. 
We implement a time-dependent protocol by connecting a series of buffers with different fuel concentrations  to the system of interest - either in a linear fashion as in Fig.~\ref{fig:baths}\,(b) or in a cycle in order to implement a cyclic protocol.

Despite the explicit apparatus, it is still possible to have implicit decision-making; a demon could move the buffers in a particular fashion following a measurement, for example. We wish to eliminate such behaviour, and make all information-processing and decision-making explicit in the system of interest, thereby avoiding conceptual pitfalls. We therefore demand that any applied protocol incorporates no external feedback based on the state of the system of interest.
One way to view the ``no external feedback" criterion is that {\it a protocol applied to a single system can be trivially applied in parallel to many systems}. This idea marginalizes all costs of applying even complex protocols. In principle, if the buffers are moved slowly enough, there is no lower bound on the mechanical cost of moving them, \revb{given pistons of arbitrarily low friction}. If we act on many systems in parallel, then any external cost associated with implementing the protocol can be taken as an edge effect. Further, if the differences in fuel concentration in consecutive buffers tends to zero, so does any thermodynamic cost associated with fuel molecules being exchanged between adjacent buffers via the volume of interest. In this limit, therefore, all thermodynamic costs arise from actual chemical reactions involving the system of interest. We now illustrate systems and protocols that implement erasing, copying and Szilard's engine. 

\subsection{Erasing a bit}
\label{sec:erasing}
A molecule with two long-lived states is a natural analog of a bit (\cite{Ouldridge_comput_2017}). Let $X$ and $X^*$ be two states of a molecule, and let us assume that it is possible to switch between $X$ and $X^*$ by coupling to fuel molecules $F_1$ and $F_1^*$, and a catalyst $Y$:
\begin{equation}
X + F_1^* + Y \rightleftharpoons X^* +F_1 +Y.
\label{eq:X_X*}
\end{equation}
The catalyst $Y$ must be present for the reaction to proceed at an appreciable rate, but is not consumed by the reaction. Reactions of this type are extremely common in natural systems (for example, $Y$ could be a kinase, $X$ its substrate, $F_1^*$, ATP and $F_1$, ADP). They can also be engineered as nucleic acid strand displacement reactions  (\cite{Qian2011b,Chen2013,Srinivas2017}). 
The chemical free energy change associated with the forwards reaction in Eq.~\ref{eq:X_X*} is
\begin{equation}
\begin{array}{c}
\Delta f_{\rm chem}(X + F_1^* + Y \rightarrow X^* +F_1 +Y) =  \Delta \mu_{F_1^* \rightarrow F_1} + \Delta \mu^0_{X \rightarrow X^*}
\vspace{2mm}
\\
 =\mu^0_{F_1} - \mu^0_{F_1^*} + k_{\rm B}T \ln \frac{[F_1]}{[F_1^*]} +  \mu^0_{X^*} - \mu^0_{X}.
\label{eq:fchem}
\end{array}
\end{equation}
Here,  $\Delta \mu^0_{X \rightarrow X^*}$ is the intrinsic stability difference between the two states (the $\ln [X]/[X^*]$ term is absent because we have a single molecule of $X/X^*$). For simplicity, we shall assume that $\Delta \mu^0_{X \rightarrow X^*}=0$, equivalent to assuming the bit is symmetric, as is typical. In this case, all changes in chemical free energy arise from the fuel molecules. The fundamental conclusions of this chapter are not affected by this assumption.

\begin{figure}[t]
  \includegraphics[width=0.7\textwidth]{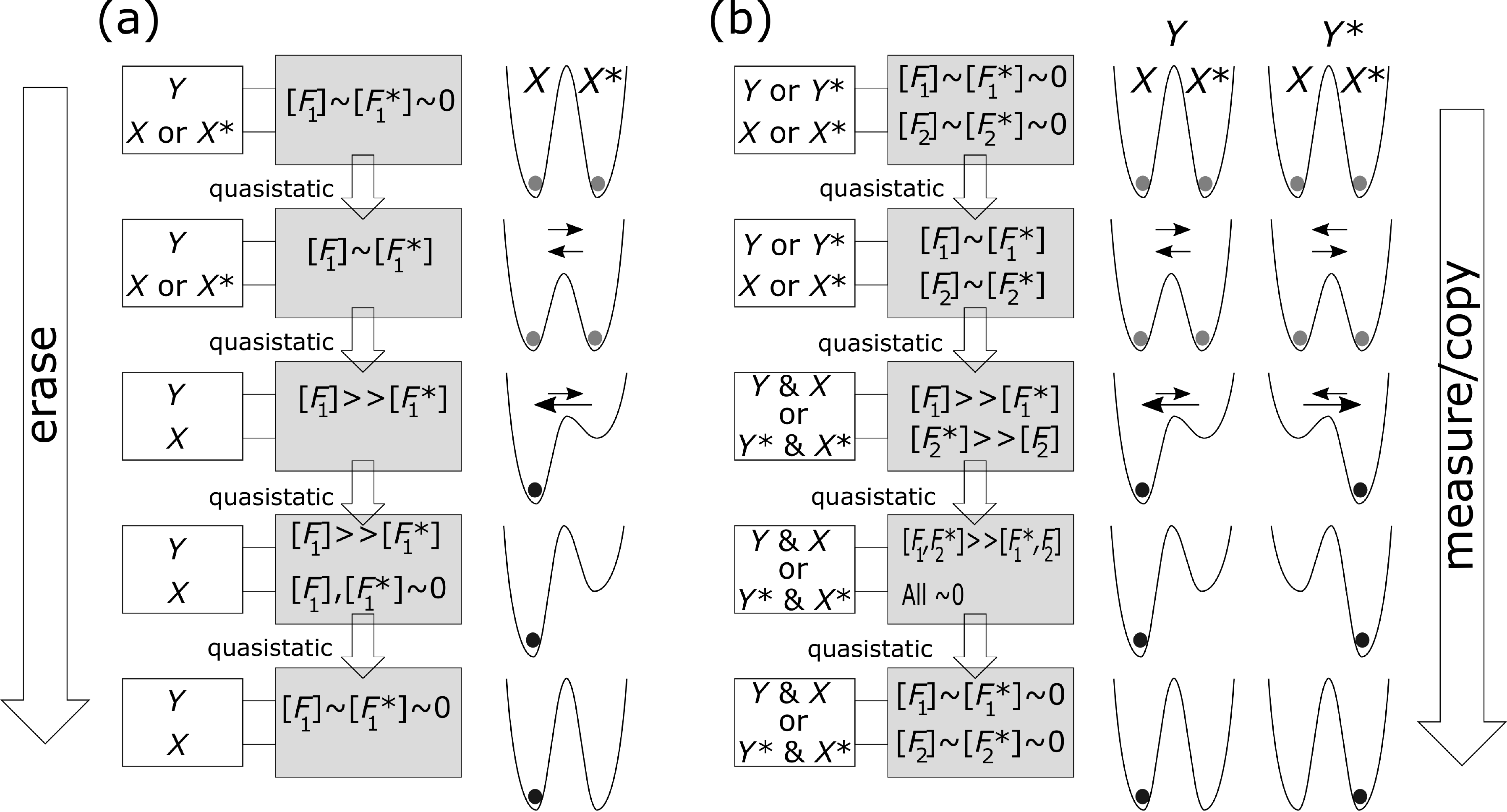}
\caption{Reversible erase and copy protocols can be implemented in molecualr systems. (a) Erasure of a molecular bit $X/X^*$ to the $X$ state. The system of intereset (far left), containing $X/X^*$ and catalyst $Y$, is coupled to a series of buffers with varying concentrations of fuel molecules; we indicate key stages in this process. $[F_1]\sim [F_1^*]$ means that $\Delta \mu_{F_1^* \rightarrow F_1}=0$; $[F] \sim 0$ means that the concentration is so low reactions requiring $[F]$ are negligible. The landscapes on the right hand represent the effective free energy landscapes for the interconversion of $X$ and $X^*$ at a given buffer and coupled to the catalyst $Y$, using an ideal reaction coordinate. Arrows indicate the dominant reaction, if any, and shaded balls indicate the probability of occupying a given state. (b) As in (a), but for copying of a $Y/Y^*$ bit using two pairs of fuel molecules. Landscapes are now shown for both possibilities of the data bit $Y/Y^*$.}
\label{fig:erase_copy}       
\end{figure}

We now implement erasure on $X/X^*$ using the apparatus outlined in Fig.~\ref{fig:erase_copy}\,(a) (\cite{Ouldridge_comput_2017}). Let a single $Y$ molecule and a single $X/X^*$ be confined in the volume. Initially, we start with buffers with both $[F_1], [F_1^*] \rightarrow 0$, so both $X$ and $X^*$ are essentially stable. We then slowly increase both $[F_1]$ and $[F_1^*]$ to a point where the forwards and backwards reactions occur at appreciable rates, maintaining a ratio of $[F_1]/[F_1^*]$ that ensures that $\Delta \mu_{F_1^* \rightarrow F_1} =0$. We now see interchanges of $X$ and $X^*$, but do not yet bias the bit's state. In the energy landscape representation  shown in Fig.~\ref{fig:erase_copy}, this initial manipulation corresponds to lowering the barrier between macrostates prior to quasistatically destabilizing one with respect to the other. Next, we slowly increase $[F_1]$ whilst keeping $[F_1^*]$ fixed, thereby increasing $\Delta \mu_{F_1^* \rightarrow F_1}$. A large and positive value of $\Delta \mu_{F_1^* \rightarrow F_1}$ pushes $X^*$ towards $X$. When we have sufficiently large $\mu_{F^*_1} - \mu_{F_1} \gg k_{\rm B}T$ that the probability of finding $X^*$ is essentially zero, we  start to decrease both $[F_1]$ and $[F_1^*]$ towards zero, whilst maintaining a constant $\mu_{F^*_1} - \mu_{F_1} \gg k_{\rm B}T$, raising the barrier and ``freezing" our molecular bit in the $X$ state. Finally, we reduce $[F_1]$ further so that $[F_1], [F_1^*] \rightarrow 0$ and $\mu_X = \mu_{X^*}$. The original buffer is now re-connected, but the initially uncertain bit has been set to a guaranteed $X$ state. 

Overall, the free energy change of the reaction volume itself is $\Delta \mathcal{F}^{\rm sys} = k_{\rm B}T \ln2$, since both $X$ and $X^*$ have the same chemical free energy, but the entropy $\mathcal{S}^{\rm sys}[p(m)] =- k_{\rm B} \sum_m p(m) \ln p(m)$ decreases from $k_{\rm B}T\ln2$ to 0.

The cost of erasure is supplied by the buffers. Initially, when $[F_1]$ and $[F_1^*]$ are increased simultaneously whilst maintaining 
$\mu_{F_1} = \mu_{F_1^*}$, no reactions occur on average. 
 Subsequently, $[F_1]$ is increased relative to $[F_1^*]$, and net reactions do occur, consuming chemical free energy and performing chemical work. If the buffers are updated  sufficiently slowly, then the $X/X^*$ system reaches a new buffer-dependent equilibrium at each point (\cite{Ouldridge_comput_2017}). From Eqs.~\ref{eq:free_energy_m} and \ref{eq:fchem}, the equilibrium probability of observing $X$ in each buffer is  
\begin{equation}
p(X) = \frac{ 1}{1+  \exp(-\Delta \mu_{F_1^* \rightarrow F_1}/k_{\rm B}T)}.
\end{equation}
When the system is coupled to the next buffer, there is an infinitesimal change ${\rm d} \Delta \mu_{F_1^* \rightarrow F_1}$. Consequently, $p(X)$ changes by 
\begin{equation}
\begin{array}{c}
{\rm d}p(X) =  \frac{ 1}{1+  \exp(-(\Delta \mu_{F_1^* \rightarrow F_1}+{\rm d} \Delta \mu_{F_1^* \rightarrow F_1})/k_{\rm B}T)} 
 -\frac{ 1}{1+  \exp(-\Delta \mu_{F_1^* \rightarrow F_1}/k_{\rm B}T)}
\vspace{2mm}
\\
= \frac{ \exp(-\Delta \mu_{F_1^* \rightarrow F_1} /k_{\rm B}T)}{\left(1+  \exp(-\Delta \mu_{F_1^* \rightarrow F_1}/k_{\rm B}T)\right)^2} \frac{{\rm d} \Delta \mu_{F_1^* \rightarrow F_1}}{k_{\rm B}T}.
\end{array}
\end{equation}

When $X^*$ changes to $X$, $F_1$ changes to $F_1^*$ in the buffer, with an associated $\Delta f_{\rm chem}^{\rm buffers}   = -\Delta \mu_{F_1^* \rightarrow F_1}$. The expected drop in free energy, or chemical work done, by the change ${\rm d}p(X)$ is
\begin{equation}
\begin{array}{c}
{\rm d} \mathcal{W}^{\rm in}_{\rm chem} = \Delta \mu_{F_1^* \rightarrow F_1} {\rm d} p(X) 
=\Delta \mu_{F_1^* \rightarrow F_1} \frac{ \exp(-\Delta \mu_{F_1^* \rightarrow F_1} /k_{\rm B}T)}{\left(1+  \exp(-\Delta \mu_{F_1^* \rightarrow F_1}/k_{\rm B}T)\right)^2} \frac{{\rm d} \Delta \mu_{F_1^* \rightarrow F_1}}{k_{\rm B}T}.
\end{array}
\end{equation}
To find the total expected work done by the all of the buffers, we simply integrate ove the whole protocol: $\Delta \mu_{F_1^* \rightarrow F_1}=0$ to $\Delta \mu_{F_1^* \rightarrow F_1} \rightarrow \infty$. Using the substitution $u =  \Delta \mu_{F_1^* \rightarrow F_1} /k_{\rm B}T$, we obtain
\begin{equation}
\begin{array}{c}
\frac{\mathcal{W}^{\rm in}_{\rm chem}}{k_{\rm B} T}
= \int_0^\infty \frac{u\exp(-u)}{(1+\exp(-u))^2} {\rm d} u
=   \int_0^\infty -\frac{\exp(u)}{1+\exp(u)} + \frac{\rm d}{{\rm d}u} \left( \frac{u\exp(u)}{1+\exp(u)} \right) \,\,{\rm d}u
=  \ln 2.
\label{eq:integral}
\end{array}
\end{equation}
The chemical work done to erase our molecular bit perfectly is $k_{\rm B} T \ln 2$, as expected. The process is thermodynamically reversible; applying the buffers in reverse would return the system to the initial state, and the overall change in the non-equilibrium free energy is zero ($W^{\rm in}_{\rm chem} = \Delta \mathcal{F}^{\rm sys}$). 

\subsection{Copying or measuring a bit}
\label{sec:copying}
To allow for copying, we consider two states of the catalyst, $Y$ and $Y^*$. We shall label the catalyst as the "data" molecule or data bit. Moreover, we shall require that $Y$ and $Y^*$ separately catalyse reactions that couple the reactions of the measurement molecule or memory bit $X \rightleftharpoons X^*$  to two distinct fuels (\cite{Ouldridge_comput_2017}):
\begin{equation}
\begin{array}{c}
X + F_1^* + Y \rightleftharpoons X^* +F_1 +Y,
\hspace{5mm}
X + F_2^* + Y^* \rightleftharpoons X^* +F_2 +Y^*.
\label{eq:reacts_copy}
\end{array}
\end{equation}
Such a scheme approximates natural "bifunctional kinases" (\cite{Stock:2000ve}), and could easily be engineered with nucleic acid circuitry  (\cite{Qian2011b,Chen2013,Srinivas2017}). 

Performing a perfect copy corresponds to correlating the data and measurement molecules.   If we knew for certain that the data molecule was in the $Y$ state, we could simply repeat the protocol outlined in Section~\ref{sec:erasing}. However, we also need the system to automatically  handle the possibility that the data molecule is in the $Y^*$ state via a single protocol.
If we follow the protocol of Section~\ref{sec:erasing}  and Fig.~\ref{fig:erase_copy}\,(a), but with the data molecule in the $Y^*$ state, nothing would happen, because reactions require the presence of either $F_1$, $F_1^*$ and $Y$ or $F_2$, $F_2^*$ and $Y^*$. We can therefore  implement two erasures simultaneously as a single protocol:  driving the system to $X^*$ if $Y^*$ is present, and to $X$ if $Y$ is present. Such a protocol is illustrated in Fig.~\ref{fig:erase_copy}\,(b). The buffer concentrations of $[F_2]$ and $[F_2^*]$ mirror those of $[F_1]$ and $[F_1^*]$, in that $[F_2]$ and $[F_2^*]$ are initially increased at $\Delta \mu_{F_2^* \rightarrow F_2}=0$, but then $[F_2^*]$ is increased further so that 
$-\Delta \mu_{F_2^* \rightarrow F_2} \gg k_{\rm B}T$, and the system is driven to towards the $X^*$ state if $Y^*$ is present. 

Calculating the chemical work done during the copy is trivial. We assume for simplicity that there is a 50\% chance that $Y$ is present, and a 50\% chance that $Y^*$ (similar arguments can be constructed for any initial bias). In either case, the decrease in chemical free energies of the buffers is $k_{B}T \ln 2$, and so $\mathcal{W}^{\rm in}_{\rm chem} = k_{B}T \ln 2$.
This decrease in free energy exactly matches the increase of $\mathcal{F}^{\rm sys}$. As in Section~\ref{sec:erasing}, the chemical free energy of the system of interest is unchanged, but $ \mathcal{S}[p(m)]$ has decreased from $k_{\rm B}\ln 4$ to $k_{\rm B}\ln 2$  -- we have gone from four equally probable macrostates to just two due to the correlations. We have paid chemical work to decrease the entropy of the data/measurement molecule system, thereby storing free energy. 

Crucially, the protocol outlined above is true measurement in the sense discussed by Szilard. The two molecules' states are correlated without direct interaction, and were the catalyst (the data bit) to undergo a state change, there would be no tendency for the measurement molecule (memory bit) to change. We have previously described this type of copying as {\it persistent} (\cite{Ouldridge_comput_2017}).

\subsection{A molecular Szilard engine}
\label{sec:Szilard}
Having implemented a molecular copy (or measurement), we now exploit the measurement through feedback. The essential challenge is to allow the data molecule $Y/Y^*$ to evolve so that it decorrelates with the fixed memory $X/X^*$, releasing the free energy stored in correlations in a controlled way so that chemical work can be extracted. Decorrelation could also be performed by undoing the measurement: here, $Y/Y^*$ does not change but $X/X^*$ is randomized. This process, which doesn't involve feedback, can be achieved reversibly by reversing the copy protocol.

Feedback is normally implicit in Szilard's engine; it is difficult to imagine concrete systems in which one degree of freedom influences the other during an initial period, and then this influence is reversed. Fortunately, our chemical system allows this behaviour. On top of the reactions used during the copy (Eq.~\ref{eq:reacts_copy} ), we consider catalytic interconversion of $Y/Y^*$ by $X/X^*$ using the fuel molecules. $G_1$, $G_2$, $G_1^*$ and $G_2^*$:
\begin{equation}
\begin{array}{c}
Y + G_1^* + X \rightleftharpoons Y^* +G_1 +X, \hspace{5mm} 
Y + G_2^* + X^* \rightleftharpoons Y^* +G_2 +X^*.
\end{array}
\end{equation}
The system is now a mutual bi-functional catalyst network. Such a system is unusual, but by no means unimaginable.

\begin{figure}[t]
  \includegraphics[width=0.85\textwidth]{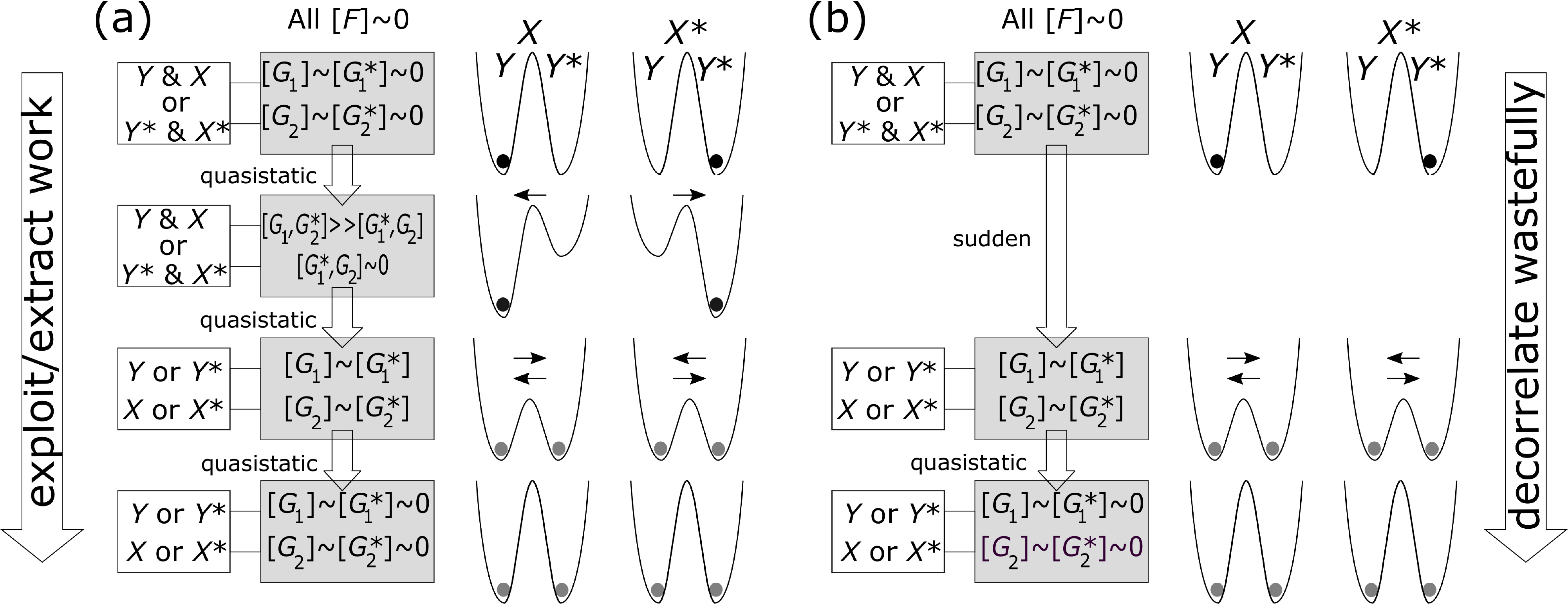}
\caption{The work extraction step of a Szilard engine as implemented by a molecular system. (a) Work extraction step following the copy/measurement step in Fig~\ref{fig:erase_copy}\,(b). The final state is equivalent to the initial state of the measurement protocol, so combining the two produces a full cyclic operation. The combined data and measurement molecules are exposed to a series of buffers of varying fuel concentrations; key points are illustrated. $[G_1]\sim [G_1^*]$ means that $\Delta \mu_{G_1^* \rightarrow G_1}=0$; $[G] \sim 0$ means that the concentration is so low reactions requiring $[G]$ are negligible. 
The landscapes on the right hand represent the effective free energy landscapes for the interconversion of $Y$ and $Y^*$ at a given buffer and coupled to either  $X$ of $X^*$ acting as a catalyst. Arrows indicate the dominant reaction, if any, and shaded balls indicate the probability of occupying a given state. (b) Relaxation of the correlated state without careful extraction of chemical work during the decorrelation. Unlike in the scheme of (a), work is not extracted by slowly changing the relative stability of the two states.}
\label{fig:molecular_szilard}       
\end{figure}

The overall protocol of our molecular Szilard engine is illustrated in Fig.~\ref{fig:molecular_szilard} \,(a). The initial measurement, or correlation stage, proceeds exactly as in Section~\ref{sec:copying}. The concentration of fuel molecules $[G_1], [G_1^*], [G_2], [G_2^*] \rightarrow 0$, preventing the state of the data molecule from changing during this period. This part of the protocol requires $\mathcal{W}^{\rm in}_{\rm chem} = k_{B}T \ln 2$.

The subsequent exploitation stage proceeds as follows. First, $[G_1]$ and  $[G_2^*]$  are increased whilst maintaining $[F_1], [F_1^*], [F_2],[F_2^*], [G_1^*], [G_2] \rightarrow 0$, until $\Delta \mu_{G_1^* \rightarrow G_1}, -\Delta \mu_{G_2^* \rightarrow G_2} \gg k_{\rm B}T$. Since only $(Y,X)$ and $(Y^*,X^*)$ combinations are possible at this point, no reactions occur and no chemical work is done. Subsequently, $[G_1^*]$ and  $[G_2]$ are slowly increased until $\Delta \mu_{G_1^* \rightarrow G_1} =\Delta \mu_{G_2^* \rightarrow G_2} =0$. 
During this period, the data is decorrelated from the measurement, and free energy is transferred back to the buffers. Given a measurement $X$, the data molecule is in state $Y^*$ with probability 
\begin{equation}
p(Y^*|X) = \frac{ \exp(-\Delta \mu_{G_1^* \rightarrow G_1} /k_{\rm B}T)}{1+  \exp(-\Delta \mu_{G_1^* \rightarrow G_1}/k_{\rm B}T)}
\end{equation}
when coupled to a buffer at $\Delta \mu_{G_1^* \rightarrow G_1}$. Proceeding as in Section~\ref{sec:erasing}, we multiply the increase in $p(Y^*|X)$ associated with a buffer increment of ${\rm d} \Delta \mu_{G_1^* \rightarrow G_1}$ by the resulting free energy change of the buffer for a conversion of $Y$ into $Y^*$, $\Delta \mu_{G_1^* \rightarrow G_1}$, and integrate:
\begin{equation}
\Delta f_{\rm chem}^{\rm buffers} = - \int_\infty^0  \Delta \mu_{G_1^* \rightarrow G_1}  \frac{ \exp(-\Delta \mu_{G_1^* \rightarrow G_1} /k_{\rm B}T)}{\left(1+  \exp(-\Delta \mu_{G_1^* \rightarrow G_1}/k_{\rm B}T)\right)^2}  \frac{{\rm d} \Delta \mu_{G_1^* \rightarrow G_1}}{k_{\rm B}T}.
\end{equation}
The integral is identical to Eq.~\ref{eq:integral}, and ths $\Delta f_{\rm chem}^{\rm buffers}  = k_BT \ln2$. The same $\Delta f_{\rm chem}^{\rm buffers}$ is obtained for a measurement result of $X^*$, although it is the combined free energy of the $G_2/G_2^*$ fuels that increases. Thus the average increase in the total chemical free energies of the buffers is $ -\mathcal{W}^{\rm in}_{\rm chem} = k_BT \ln2$, and so the  chemical work done in performing the measurement is extracted. Finally, no chemical work is done as the concentrations of all fuels are taken to zero at fixed $\Delta \mu_{G_1^* \rightarrow G_1} = \Delta \mu_{G_2^* \rightarrow G_2} = 0$. This final operation corresponds to "raising the barrier" between the metastable states of the bit in the energy landscape analog, as shown in Fig.~\ref{fig:molecular_szilard}.

Our molecular device is an exact analogue of the engine proposed by Szilard, with all components explicit. We perform a binary measurement by creating a record of the data molecule with the measurement molecule. Chemical work of $k_{B}T \ln 2$ is required by the measurement step, in order to create a low-entropy, high-free energy correlated state. This work is then extracted from these correlations during exploitation, when the measurement molecule itself couples the data molecule to the appropriate fuel reservoir. Importantly, the measurement is persistent in that the $X/X^*$ state does not change during exploitation.



\section{What we learn from explicit biochemical protocols}
\label{sec:gain}
\subsection{Erasure is not the most compelling explanation for the self-consistency of the second law}
\label{sec:no erase}
There is no erasure in the protocol above. In principle, we could add one by including a third molecule $Z$, which couples the transition of $X \rightleftharpoons X^*$ to another pair of fuel molecules $H$ and $H^*$. By implementing a protocol equivalent to that in Section~\ref{sec:erasing} after the exploitation step, we could reversibly set the measurement molecule to state $X$, at a cost of $k_{\rm B}T \ln 2$ of chemical work. 
Having paid for the erasure, the minimal chemical work required to go from the erased state to the post-measurement, correlated state would now be zero since $\Delta \mathcal{F}^{\rm sys}=0$; in both cases there are two equally likely macrostates for the measurement/data system ($(d,m)=(0,0)$ OR (1,0) and (0,0) OR (1,1), respectively).


Although cycles involving erasure are possible, the process is not required, and so it seems slightly perverse to attribute the self-consistency of the second law in Szilard's engine to erasure (\cite{Maroney2005}). Indeed, in  the simplest implementation of measurement with an erased bit, one first "un-erases" the measurement molecule to an equal probability of $X$ and $X^*$, prior to even coupling to $Y/Y^*$ via the $F$ fuels. During this un-erase, $k_{\rm B}T \ln 2$ of chemical work is extracted, giving zero overall chemical work for measurement from an erased state when combined with the $k_BT \ln 2$ required to subsequently create correlations. Erasing then appears to be an accounting trick; we add a process to the cycle (erase and un-erase) that achieves nothing, but then incorporate the un-erase into a "measurement from an erased state" process (Fig.~\ref{fig:Szilard})) that overall has no cost. We note that Bennett's analysis also includes an "un-erase" prior to correlation (\cite{Bennett1982,Bennett2003}). It is possible to incorporate erasure in a less trivial manner, but we still find the emphasis on any one step to be misleading. The second law is not violated at {\it any} stage of the protocol, whether erasure is performed or not (\cite{Fahn1996,Maroney2005,Sagawa2014,Parrondo2015}).

Instead, it is more illuminating to focus on the non-equilibrium free energy. As originally pointed out by Szilard, the measured state contains correlations between degrees of freedom that must persist despite an absence of direct interactions. 
The existence of correlations without direct interactions implies a low entropy state without a compensating low chemical free energy (\cite{Fahn1996,Sagawa2014,Parrondo2015}). In this case, the 
correlations mean only two of a total of four equally stable macrostates are possible; the entropy of the post-measurement state is $k_{\rm B}\ln 2$ rather than $k_{\rm B} \ln4$ and an excess free energy of $\mathcal{F}^{\rm sys}-\mathcal{F}^{\rm sys}_{\rm eq} =   k_{\rm B}T \ln 2$ is available. The stored free energy in the measured state makes the exploitation step consistent with the second law, regardless of the details of the cycle. The only difference between protocols with and without erasure is whether we chose to compensate for the creation of the measurement by initially creating an out-of-equilibrium memory bit.


This explanation of Szilard's engine generalizes. In information theory, correlations between two degrees of freedom $m_1$ and $m_2$ are described using the mutual information (\cite{Shannon1949})
\begin{equation}
\mathcal{I}(m_1, m_2) = \mathcal{I}[p(m_1), p(m_2)] = \sum_{m_1,m_2} = p(m_1,m_2) \ln \left( \frac{p(m_1,m_2)}{p(m_1)p(m_2)} \right) \geq 0,
\end{equation}
with the equality holding if the two degrees of freedom are independent. It is relatively straightforward to show that if the energy/chemical free energy is additive in $m_1$ and $m_2$ (the two degrees of freedom do not interact), then (\cite{Esposito2011,Horowitz2013,Parrondo2015})
\begin{equation}
\mathcal{F}[p(m_1), p(m_2)] =  \mathcal{F}[p(m_1)] + \mathcal{F}[p(m_2)] + k_{\rm B}T \mathcal{I}[p(m_1), p(m_2)].
\end{equation}
Since $m_1$ and $m_2$ must be independent in equilibrium, a non-zero $\mathcal{I}[p(m_1), p(m_2)] $ necessarily represents a store of free energy. This principle has been applied to systems that create and exploit correlations between non-interacting degrees of freedom (\cite{Horowitz2013,Ouldridge_comput_2017,McGrath2017,Ouldridge_polymer_2017,Brittain2018,Stopnitzky2018,Chapman2015,Boyd2016}).

Bennett's argument that, in a setting with heat and work rexervoirs coupled to a pair of bits, measurement can be performed with no work input using an erased memory, and that erasing memories requires work input, is not incorrect. We nonetheless believe that the emphasis on erasure causes unnecessary confusion. Taken at face value, the Bennett explanation might seem to suggest that the second law is transiently violated, since erasure compensates for the work extraction after the fact, when a whole cycle is completed (\cite{Landauer1991}). Indeed, the finite size of the demon's memory is sometimes invoked (\cite{feynman1998feynman,Deffner2015}),  suggesting that the second law can be violated until the memory is full and needs to be erased. But if the second law can be violated until memory has to be erased, then what happens if erasure never happens? If, instead, the necessarily non-equilibrium nature of the post-measurement state is used to support the second law, these confusions evaporate; all sub-steps are consistent with the second law (\cite{Parrondo2015,Sagawa2014}). Furthermore, the perceived importance of erasure leads researchers to look for its presence in systems that could perhaps be more fruitfully analysed in terms of the generation and usage of non-equilibrium correlations (\cite{Mehta2012,Andrieux2013}). 

\subsection{Focussing on heat and work can be misleading}
\label{sec:focus}
There is a  tendency to use heat generation as a shorthand for irreversibility, or the increase in the total entropy. This shorthand, however, can lead to significant confusion. 
Heat generation does not imply thermodynamic irreversibility. It is  possible, at least in principle, to transfer heat to a system's environment in a thermodynamically reversible way, provided that the increase in the entropy of the environment is compensated by a decrease in the entropy of the system. A common example would be quasistatic compression of a piston.  Heat transfer to the environment only guarantees irreversibility if the system operates cyclically, in which case its entropy cannot change to compensate. Nonetheless, the transfer of heat is frequently described as "dissipation", without clarifying the degree to which the \revb{total entropy} actually increases (\cite{Maroney2005}). 

\revb{From a technical perspective, the distinction is between entropy transferred and entropy produced. The entropy production is the increase in total entropy. It is the sum of the change in entropy of the system of interest and the entropy transferred to the rest of the environment due to its interaction with the system. In the case where the system of interest is in contact with a single heat reservoir the entropy transferred is the heat transferred to the reservoir divided by the temperature (\cite{Parrondo2015}) -- the heat transferred is therefore not directly informative of the total entropy production, even in simple settings settings with a single heat reservoir.}

In Landauer's analysis of erasure, heat generation compensates for the entropy decrease of the bit. Despite the fact that the bound of $k_{\rm B}T \ln 2$ is explicitly derived for a reversible protocol satisfying $\Delta \mathcal{S}^{\rm tot} = 0$, \revb{Landauer himself describes the heat transfer as a necessarily thermodynamically irreversible process (\cite{Landauer1961}). Many others have apparently repeated this logic (\cite{Plenio2001,Bub2001,Ladyman2007,Dillenschneider2009,berut2012,Jun2014,Hong2016}). This confusion contributes to the erroneous narrative that thermodynamic irreversibility of erasure pays for a transient violation of the second law during a cycle of Szilard's engine, and leads to statements such as  "logical irreversibility is associated with physical irreversibility  and requires a minimal heat generation" (\cite{Landauer1961}).} 

\reva{We defined thermodynamic reversibility in Section~\ref{sec:reversibility}; it is a statement about whether a particular protocol that converts a system's distribution from $p(x)$ into $p^\prime(x)$ would recreate $p(x)$ when applied in a time reversed fashion to an initial distribution $p^\prime(x)$. Thermodynamic reversibility depends on the actual physical process undergone to move from $p(x)$ to $p^\prime(x)$. However,  exact trajectories $x (t)$ do not need to be reversed; $p(x)$ need only be re-created in the statistical sense. Whether or not a process is thermodynamically reversible depends upon both the protocol and the initial distribution $p(x)$ (\cite{Kolchinsky2017}); in general, a protocol that is thermodynamically reversible for one $p(x)$ will not be reversible for another.}

\reva{
By contrast, a process is logically reversible if and only if the initial  logical state $y_i$ can be inferred unambiguously from the final logical state $y_f$---which is not true for erasure, for example \cite{Sagawa2014}. Logical irreversibility is then solely a condition on the overall transition matrix between input and output logical states $T_{y_f y_i}$ and is independent of the initial state to which $T_{y_f y_i}$ is applied; the possibility of multiple physical states $x$ within a logical state $y$; and the detailed physical mechanism by which $T_{y_f y_i}$ is achieved. Logical reversibility is, however, crucially dependent on the specific input/output map represented by $T_{y_f y_i}$ rather than just the overall statistics. Given these differences between logical and thermodynamic irreversibility, it is not surprising that there is in fact no causal connection between the two; both logically reversible and logically irreversible processes can in principle be implemented in thermodynamically reversible and irreversible ways. \cite{Sagawa2014} has given specific examples to illustrate this fact in general settings.}

Moreover, if a process as a whole is thermodynamically irreversible, one cannot immediately associate the part of the process that transfers heat to the environment (or requires work input) as the cause of the irreversibility. For example, irreversibility in certain systems has been ascribed to the need to erase (\cite{Mehta2012,Mehta2016,Andrieux2013}). But if the erasure considered by Landauer is not necessarily thermodynamically irreversible, then it is a logical non-sequitur to attribute thermodynamic irreversibility to the presence of erasure. Landauer's principle in isolation cannot explain an  increase in the \revb{total entropy of a closed system}.
For example, consider the molecular Szilard engine including erasure as outlined in Section~\ref{sec:no erase}. This device is thermodynamically reversible. We could, however, consider an alternative protocol in which, instead of carefully relaxing the correlated state after the measurement, we simply suddenly expose the system to buffers with high concentrations of $[G_1]$, $[G_1^*]$,  $[G_2]$ and  $[G_2^*]$ with $\Delta \mu_{G_1^* \rightarrow G_1} =\Delta \mu_{G_2^* \rightarrow G_2} =0$, as in Fig.~\ref{fig:molecular_szilard}\,(b), prior to erasing. The measurement and data molecules would immediately decorrelate, and no chemical work would be extracted. The overall cycle would then require $\mathcal{W}_{\rm chem} = k_{\rm B} T \ln 2$, the total free energy of the buffers would decrease in a cycle, and the total entropy would increase. The same uncontrolled decorrelation would happen if we tried to erase directly from the correlated state using $Z$ and fuels $H$ and $H^*$, without gradually undoing the correlations first.

It is tempting to associate the irreversibility in this cycle with a step at which we have to expend net chemical work due to Landauer's principle -- erasure. In the setting of magnetic bits, this process would require work input and generate heat. However, in either context, erasure  is thermodynamically reversible. The (chemical) work input is exactly compensated by the increase in $\mathcal{F}^{\rm sys}$. Rather, it is the decorrelation step -- which requires no (chemical) work, and exchanges no heat with the environment in either the molecular or magnetic contexts -- that is the {\it cause} of the irreversibility (\cite{Fahn1996,Ouldridge_comput_2017}). Reversing this step would not restore the correlated state, and during this step $\mathcal{F}^{\rm sys}$ decreases by $k_{\rm B}T \ln 2$ due to the loss of correlations, without a compensating extraction of (chemical) work, implying an increase in the \revb{combined entropy of the system and environment}. 
 
The above discussion highlights a further issue; the second law doesn't only apply to the mechanical heat engines that drove the industrial revolution; it also applies to the molecular systems that underlie life. In these settings, and others, mechanical work and heat can play a less direct role, with processes relying on other forms of free energy transduction. Indeed, the erasure, copying and Szilard engine protocols described in Section~\ref{sec:molecular demon} involve no mechanical work at all.  
This distinction matters  because, \reva{as discussed in Section~\ref{sec:equilibrium}, free energy is not an energy; it is not conserved. If {\it chemical} work $ W^{\rm in}_{\rm chem}$  is done on a system during a cyclic evolution (in which $\Delta U =0$), physical heat  $Q^{\rm out} = W^{\rm in}_{\rm chem} $ need not be transferred to the environment to compensate.} Nonetheless, heat generation is often used as a shorthand for thermodynamic irreversibility,  \revb{even in molecular contexts (see e.g. \cite{Lan2012,Mehta2016,govern2014,Barato2017}),} as noted in \cite{Seifert2011}. Heat flow is a real energy flux that can be measured; it is misleading to describe entropy generation in a different form in this way. For example, one might incorrectly assume that a living system needs to develop machinery to rid itself of heat exactly equal to $T\dot{\mathcal{S}}$, where $\dot{\mathcal{S}}$ is its entropy generation rate, to maintain a constant temperature.

\subsection{The benefits of having an explicit supply of chemical work}
\label{sec:explicit buffer}
Work reservoirs are not usually explicitly modelled like the chemical work buffers in Section~\ref{sec:molecular demon}. Using abstract (and interchangeable) work buffers is often conceptually useful,  but it can help to make certain results seem overly remarkable. For example, measurement can be performed  without net (chemical) work expenditure if the memory is initially in a well-defined state, suggesting that the measurement has no cost. Given that the correlated post-measurement state stores free energy that can be extracted as work, this lack of a cost seems remarkable. However, there is a cost -- the initial non-equilibrium state of the memory is consumed to pay for the measurement.  At a fundamental physical level, there is no real difference between using free energy stored in the memory bit or a (chemical) work buffer. Either way, a resource is consumed to create non-equilibrium correlations between memory and data. There is no reason to value resources in a (chemical) work reservoir over and above sources of free energy explicitly incorporated into that system of interest, which is particularly clear when the work reservoir is made explicit. 

An explicit supply of (chemical) work also helps establish what is definitely possible -- a so-called "positive result". Work reservoirs are often pictured as masses in a gravitational field, as in Fig.~\ref{fig:Szilard}. It is often imagined, however, that work can be done by implementing an arbitrary change in the potential energy of a system as a function of its internal coordinates (eg. \cite{Schmiedl2007}), despite the fact that a single weight would actually be insufficient even for the simple Szilard engine of Fig.\ref{fig:Szilard}. Without an explicit mechanism to achieve the required work coupling, it is arguable whether one can positively prove the possibility of certain operations  -- even if one can prove the impossibility of others. Perhaps even more challenging than implementing an arbitrary control of a potential is the idea of recovering work back from the system of interest, as is required by Szilard's engine (\cite{Deshpande2017}). 

The operations that can be achieved with the molecular setup outlined in Section~\ref{sec:molecular demon} are clearly limited. However, they are at least definitely possible, albeit  in a highly-idealised limit. Furthermore, since the interconversion of fuel species directly determines the chemical work done, any chemical work recovered is indeed stored. By contrast, consider recent experiments in which devices such as optical feedback traps are used to implement an arbitrary force on a colloid (\cite{berut2012,Jun2014}). Although this force is designed to mimic a conservative, time-dependent potential, the thermodynamic cost of applying the control is orders of magnitude higher than calculated, and work is not actually recovered when the colloid moves against the pseudo-potential.  

\section{Biochemical systems as a platform for studying fully autonomous systems}
\label{sec:autonomy}

Although conceptually useful, the apparatus  in Section~\ref{sec:molecular demon} is highly idealised. We assume that molecular states are stable unless enzymes are present, and that the fuels only couple to the intended enzymes (in practice, lifetimes need to be long and  leak rates need to be slow relative to intended reactions). Furthermore, although we did not consider intelligent intervention or implicit decision-making, it is still somewhat unsatisfying to invoke an external protocol at all. The most intellectually-satisfying description would include the device (or demon) implementing any protocol as part of the system itself (\cite{Mehta2016}).

Such systems, which do not involve an externally-applied time-varying control, are {\it autonomous}. Molecular systems are even more suited to exploring autonomous contexts than those with externally-applied protocols. Fixed buffers of non-equilibrium fuel concentrations can drive active  processes. Crucially, diffusion is automatically present in molecular systems, allowing an arbitrary complex network of interactions at no cost. We have previously shown that a simple, autonomously-operating enzymatic network can be mapped directly to a copy process in the sense of Szilard's engine (\cite{Ouldridge_comput_2017}). The network could not reach the bound set by the non-autonomous, quasistatic protocols considered in Section~\ref{sec:molecular demon}. It could come remarkably close, however, despite operating at a finite rate, and requiring less stringent assumptions about the slowness of unintended leak reactions.

\section{Conclusion}
\label{sec:conc}
Basing our discussion around molecular protocols for manipulating bits, we have demonstrated the advantages of being more explicit in our analysis of fundamental  thermodynamic and computational systems. We have argued that doing so removes much of the mystery surrounding the physics of computation and measurement, and thermodynamics more generally. We have shown that molecular realisations are a particularly helpful way to understand basic thermodynamic ideas.

We argue that it is time to revisit the generally accepted explanation for the self-consistency of the second law of thermodynamics during the operation of Maxwell's demon and Szilard's engine. Attributing the survival of the second law to one part of the process or another is an accounting exercise, and misses the fundamental point: the second law is self-consistent throughout all stages of operation. Nonetheless, Szilard's original contention that creating a correlated state is fundamental to the device, and that creating this correlated state requires "compensation", is essentially correct. Whether this cost is paid by exploiting a reservoir, or a  supply of free energy that is part of the explicitly-modelled system, is moot.  We emphasize that Bennett's analysis is not wrong; merely the significance of erasure in the context of the second law has since been overstated.  \revb{The lack of a relationship between logical irreversibility and thermodynamic reversibility more generally calls into question the value of pursuing logically reversible computing approaches in an effort to reduce power consumption (\cite{Grochow2018}), although tracking the amount of erasure in a human-designed computations is potentially useful for some specific approaches to low-cost computation.} We should also not impose a language of erasure on (biological) systems for which it is unnatural.

\reva{We are not claiming to advance this view for the first time in this chapter, and we have drawn upon a number of works published throughout the last few decades (including \cite{Fahn1996,Maroney2005,Sagawa2009,Sagawa2014,Parrondo2015}). One might ask whether the development of new theories and tools since the middle of the 20th Century has led to this alternative rationalisation. On the one hand, modern methodology is not really necessary to understand the problem. Non-equilibrium generalized free energies and the language of information theory are a convenient tool that make Szilard's argument more formal, but the essential insight was already provided in 1929, and Landauer and Bennett's calculations themselves were not incorrect. Really it is a question of interpretation, and the development and application of stochastic thermodynamics (\cite{Jarzynski1997,Crooks1999,Esposito2011,Seifert2005}) has facilitated a much deeper general understanding of information and thermodynamics for small, fluctuating systems in recent years. This general understanding has in turn yielded a clearer approach to this particular problem. We note that for many (including ourselves), this insight has been developed whilst studying concrete molecular systems that accomplish information-processing or computational tasks (\cite{Ouldridge_comput_2017}).}

We additionally make the case for more careful use of the terms heat and work -- heat is not a synonym for an increase in the total entropy, and free energy is not an energy. The distinction between free energies and work touches on a further important point: fundamentally, chemical work buffers are reservoirs of free energy due to an entropic component to their deviation from equilibrium. Much has been made of the fact that a low-entropy string of erased bits is a thermodynamic resource (\cite{feynman1998feynman,Mandal2012}). However, entropic contributions to stored free energies are natural in molecular contexts. The effective force driving the diffusion of uncharged molecules from a high concentration to a low concentration is entirely entropic. Once the physical object encoding the low entropy state is made explicit, "information reservoirs" typically seem much less mysterious.  

We have also argued that molecular systems are a convenient basis in which to ground explicit (if idealised) representations of processes that we wish to analyse at a fundamental level. Explicit mechanisms can demystify the underlying concepts, and without a method of implementing a protocol, it is difficult to argue for positive rather than negative results. Rao and Esposito (\cite{Rao2018}) have introduced systematic methods for analysing more general molecular systems - although care should be taken when interpreting their free energies, which are not defined in the conventional manner discussed here. Furthermore, without a concrete mechanism, it is easy to design systems that are not even thermodynamically well-defined -- as discussed in \cite{Stopnitzky2018}. Needless to say, despite our enthusiasm with molecular contexts and their relevance to complex natural systems, alternative paradigms may be more relevant in other settings. 

Full autonomy is the natural limit of explicitness. Currently, there is a major lack of understanding of the possibilities of autonomous systems (\cite{Mehta2016}). Can they approach the performance of externally manipulated systems, particularly in the limit of finite time operation? How challenging is it to develop systems in which one subsystem effectively generates time-varying conditions for another, and is this effective? Molecular systems are both an important conceptual tool and an experimental testing-ground for exploring these ideas, particularly as we move beyond the classic questions of copying and erasing. 

\section{Acknowledgements}
We greatfully thank Dave Doty, Manoj Gopalkrishnan and Nick Jones for helpful conversations.

\section{Funding}
T. E. O. acknowledges support from a Royal Society University  Research  Fellowship , R.  A.  B.  acknowledges support from an Imperial College London AMMP studentship, and for P. R. t. W. this work is part of the research programme of the Netherlands Organisation for Scientific Research (NWO) and was performed at the research institute AMOLF.


\bibliography{Bibliography_thermo}
\bibliographystyle{unsrt}

\end{document}